\newcommand{\PRL}[3]{Phys.\ Rev.\ Lett.\ {\bf #1},\ #2 (#3)}
\newcommand{\RMP}[3]{Rev.\ Mod.\ Phys.\ {\bf #1},\ #2 (#3)}

\newcommand{\SC}[3]{Science\ {\bf #1},\ #2 (#3)}

\newcommand{\PRA}[3]{Phys.\ Rev.\ A\ {\bf #1},\ #2 (#3)}

\newcommand{\NJP}[3]{New \ J.\ Phys.\ {\bf #1},\ #2 (#3)}

\newcommand{\appsection}[1]{\let\oldthesection\thesection
  \renewcommand{\thesection}{ \oldthesection}
  \section{#1}\let\thesection\oldthesection}

\documentclass[twocolumn,showpacs,superscriptaddress,preprintnumbers,amsmath,amsfonts,amssymb,notitlepage,pra]{revtex4-1}
\usepackage{geometry}
\date{\today}
\DeclareMathAlphabet{\mathpzc}{OT1}{pzc}{m}{it}
\usepackage[utf8x]{inputenc}
\usepackage{amsmath}
\usepackage{amsfonts}
\usepackage{amssymb}
\usepackage{graphicx}
\usepackage{nicefrac}
\usepackage{amsbsy}
\usepackage{float}

\def \be{\begin{equation}}
\def \ee{\end{equation}}
\def \ba{\begin{array}}
\def \ea{\end{array}}
\def \bea{\begin{eqnarray}}
\def \eea{\end{eqnarray}}

\begin{document}
\title{Absence of the Twisted Superfluid State in a mean field model of bosons on a Honeycomb Lattice}
\author{Sayan Choudhury}
\email{sc2385@cornell.edu}
\author{Erich J Mueller}
\email{em256@cornell.edu}
\affiliation{Laboratory of Atomic and Solid State Physics, Cornell University, Ithaca, New York}
\begin{abstract}
Motivated by recent observations (P. Soltan-Panahi {\it et al.}, Nature Physics {\bf 8}, 71-75 (2012)), we study the stability of a Bose-Einstein Condensate within a spin-dependent honeycomb lattice towards  forming a ``Twisted Superfluid" state. Our exhaustive numerical search fails to find this phase, pointing to possible non-mean field physics.
 \end{abstract}

\maketitle

\section{Introduction}

\subsection{Background}

Recently Soltan-Panahi {\it et al.} found evidence of a zero quasi-momentum ``Twisted Superfluid" state of a two-component Bose-Einstein condensate (BEC) trapped in a spin-dependent honeycomb lattice \cite{Sengstocktsfnsf}.  A twisted superfluid is characterized by Bose-Einstein condensation into a state whose order parameter (a macroscopically occupied single particle wavefunction) has a spatially varying phase. The simplest example is condensation at finite momentum. Alternatively, in a non-Bravais lattice where the unit cell involves multiple sites, one can have a twisted superfluid at zero quasi-momentum if the phase of the order parameter varies throughout the unit cell. We model Soltan-Panahi et al.'s experiment \cite{Sengstocktsfnsf} with a mean field Gross-Pitaevskii function. We find that the twisted superfluid state is absent within mean field theory thus suggesting that the observations are due to non-mean field effects.\\

Twisted Superfluids are quite exotic; the phase twists of the order parameter are naturally associated with microscopic currents. Moreover, the present example involves spontaneous symmetry breaking, and provides a setting for studying phase transition physics. Analogous physics can be found in magnetic systems \cite{SengstockFrustratedScience2011} and in the excited states of lattice bosons \cite{Hemmerichpwave2011,Hemmerichfwave2011}. 

\subsection{Experimental Evidence for a Twisted Superfluid}

In their experiment \cite{Sengstocktsfnsf}, Soltan-Panahi et al. created a two component Bose-Einstein condensate (BEC)  of $\rm ^{87}Rb$ atoms in a spin-dependent honeycomb lattice.  Soltan-Panahi {\it et al.}  find evidence for the Twisted Superfluid state in two cases: a BEC of $\rm ^{87}Rb$ atoms in the $|F=1,m_F=-1\rangle$ and  $|F=1,m_F=1\rangle$ state and a BEC of $\rm ^{87}Rb$ atoms in the $|F=2,m_F=-2\rangle$ and $| F=1,m_F=-1\rangle$ state. In both of these cases, the two spin states form out-of-phase charge density waves in this spin dependent lattice. In Figure 1, we show a cartoon of the density of atoms in one of the two spin states. For the rest of this paper, we focus on the case where the two spin states are $| F=1,m_F=-1\rangle$ and  $| F=1,m_F=1\rangle$.\\

The main experimental evidence for non-trivial phases of the superfluid order parameter  comes from time-of-flight expansion, a technique where all trapping fields are removed and the atomic ensemble falls freely under gravity. Neglecting interactions \cite{KupferschmidtMueller}, the long-time real space density profile is simply the initial density in momentum space. For the special case of a BEC, the momentum space density, $n_k$ is the Fourier transform of the order parameter : $n_k = |\psi({\bf k})|^2=|\int \exp(+ i {\bf k.r}) \psi({\bf r})|^2$, where $\psi({\bf r})$ is the order parameter of the BEC. As schematically illustrated in Figure 2, if $\psi({\bf r})$ is real, and has the symmetry of the honeycomb lattice, its Fourier transform (and consequently the time-of-flight pattern) is six fold symmetric. This six-fold symmetry persists even if the densities on the two sub-lattices differ, forming a three-fold symmetric charge density wave as illustrated in Figure 1. Mathematically, this six-fold rotational symmetry of the time-of-flight pattern is a consequence the point group symmetry of the lattice ($C_{3v}$) and the relation $\psi({\bf - k})=\psi^{*}({\bf k})$, which holds for real $\psi({\bf r})$. Therefore, a time-of-flight pattern without inversion symmetry ($\psi({\bf - k}) \ne \psi^{*}({\bf k})$) is direct evidence of a complex wavefunction (i.e. a twisted superfluid state). The experimentalists see exactly this signature. \\

From the time-of-flight images obtained in \cite{Sengstocktsfnsf}, a breakdown of the six-fold rotational symmetry in momentum space is observed for lattice depths $V_{\rm lat}$ ranging from about 1 to 4 $E_{\rm R}$, where $E_{\rm R} = \frac{\hbar^2}{2 m \lambda_L^2}$, $m$ is the mass of $\rm ^{87}Rb$ atoms and $V_{\rm lat}$ is precisely defined by Eq.(6).  Figure 2 illustrates this structure in which the amplitudes of the first order time-of-flight peaks (denoted by $|t|$ and $|z|$) have different values for this range of lattice depths. An important aspect of their experiment was that this rotational symmetry breaking arises only if both species of atoms are present.  Moreover, the symmetry breaking was opposite for the two species (i.e $\frac{|t_1|}{ |z_1|} = \frac{ |z_2|}{ |t_2|}$). The order parameter (OP) for the twisted superfluid state is  given by:
\be
OP = |\frac{|z|^2-|t|^2}{|z|^2+|t|^2}|
\ee

By construction, $OP$ has a non-zero value in the twisted superfluid and is zero for a uniform condensate. Soltan-Panahi {\it et al.} measure this quantity.\\

The experimental evidence suggests that the order parameter is uniform on each of the triangular sub-lattices of the honeycomb lattice, but that there is a relative phase $\delta$ between them.  
\bea
|z|^2 &=&  n_{+} + n_{-} + 2\sqrt{n_{+}n_{-}} \sin({\delta}) \,\,\,\ \rm and \\
|t|^2 &=&  n_{+} + n_{-} - 2\sqrt{n_{+}n_{-}} \sin({\delta}),
\eea
where the $\rm n_{+}$ and $\rm n_{-}$ denote the density of atoms on the two distinct sub-lattices. Thus, the order parameter is :
\be
OP = \frac{2\sqrt{n_{+}n_{-}} |{\sin({\delta})}|}{n_{+} + n_{-} }.
\ee

\begin{figure}
\begin{center}
\includegraphics[scale=0.44]{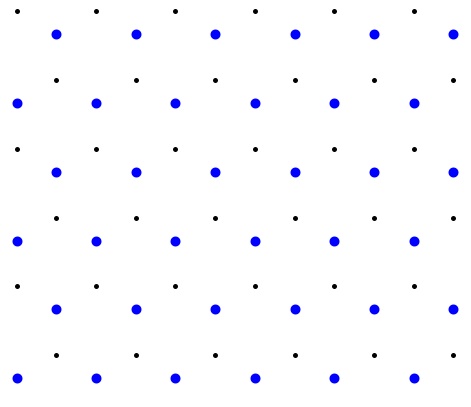}
\end{center}
\caption{The density wave formed in a honeycomb lattice for the $m_F = 1$ atoms. The points represent lattice sites. Larger points indicate a site filled with more atoms. This pattern is periodically repeated. A complementary density wave is formed by $m_F=-1$  atoms. This density wave does not lead  to a 6-fold symmetry breaking in time-of-flight unless additional phases appear on the sites.}
\label{cdw}
\end{figure}

\begin{figure}
\begin{center}
\includegraphics[scale=0.55]{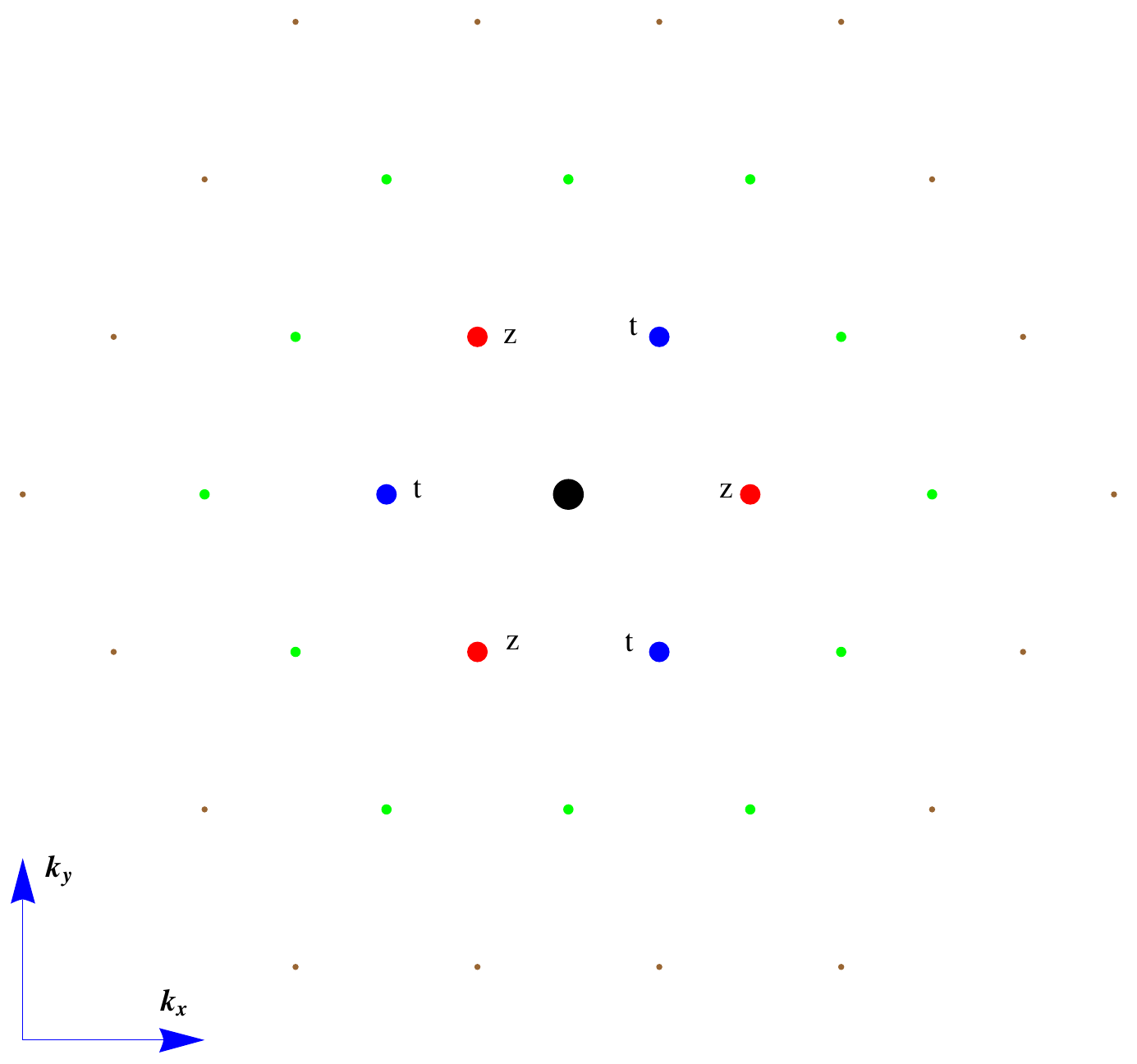}
\end{center}
\caption{Schematic of the Time-of-Flight pattern for a superfluid in a 2D honeycomb lattice. Larger darker dots correspond to more particles with a given momentum. The complex numbers $|$t$|$ and $|$z$|$ represent the amplitudes of the Fourier transform of the condensate wavefunction at $k = (\frac{\pi}{a},0)$ and $k=(\frac{\sqrt{3} \pi}{2a},\frac{\pi}{2a})$ (see text). The twisted superfluid  is described by $|t| \ne |z| $.}
\label{tsfnsf}
\end{figure}

\section{The Model}

Within a mean field model, we will investigate the relative stability of twisted or ordinary superfluids. The energy of a two component BEC,  described by macroscopic wavefunctions $\psi_1$ and $\psi_2$ is :
\bea
E_{\rm 3D} &=& \int d^3{\bf r} \sum_{\sigma=1,2}  [\frac{\hbar^2}{2m}|\nabla \psi_\sigma({\bf r})|^2  +  V_{\sigma}({\bf r}) |\psi_\sigma({\bf r})|^2 \nonumber\\ 
&+& \frac{U_{\rm 3D}^{\sigma}}{2}|\psi_\sigma({\bf r})|^4] + W_{\rm 3D} |\psi_1({\bf r})|^2  |\psi_2({\bf r})|^2   \nonumber\\
&+& V_{\rm conf}({\bf r}) (|\psi_1({\bf r})|^2 + |\psi_2({\bf r})|^2)
\label{Energy}
\eea

Here, $U^{\sigma}_{\rm 3D}=\frac{4 \pi \hbar^2 a_{\sigma}}{m}$ is the intra-species interaction energy ($a_{\sigma}$ is the intra-species scattering length for species $\sigma$), while $W_{\rm 3D}=\frac{4 \pi \hbar^2 a_{12}}{m}$ is the inter-species interaction energy ($a_{12}$ is the inter-species scattering length). As already mentioned in Section I B, we focus on the case in \cite{Sengstocktsfnsf}, where the states 1 (described by $\psi_1$) and 2 (described by $\psi_2$) are the $|F=1,m_F=1\rangle$ and $ |F=1,m_F= - 1\rangle$ states of $\rm ^{87}Rb$.  For these two hyperfine states of $^{87}\rm Rb$ atoms, $U^{1}_{\rm 3D}$,$U^{2}_{\rm 3D}$and $W_{\rm 3D}$ are almost equal ($a \approx 100 a_0$ where $a_0$ is the Bohr radius). In principle collisions can connect these hyperfine states to others (for example $|F=1,m_F=0\rangle$). For the experimental parameters, these processes are off-resonant and the two-component Bose gas model describes the physics. \\

In the experiment \cite{Sengstocktsfnsf}, the honeycomb lattice is generated by 3 lasers yielding a potential $V_i({\bf r}) = V_{\rm hex}({\bf r}) \pm \alpha B_{\rm eff} ({\bf r})$ where, state 1 sees the sign `+' and state 2 sees the sign `-' (with $\alpha = 0.13)$ and
\bea
V_{\rm hex}({\bf r}) &=& 2\,\ V_{\rm lat}( \cos[{\it k_L}\,\ {\bf b_1.x}] +  \cos[{\it k_L}\,\ {\bf b_2 .x}]   \nonumber \\ 
&+& \cos[{\it k_L}\,\ {\bf b_3.x}] )\\
  \nonumber\\
B_{\rm eff} ({\bf r}) &=& 2\sqrt{3} \,\ V_{\rm lat} ( \sin[{\it k_L} \,\ {\bf b_1.x}] + \sin[{\it k_L} \,\ {\bf b_2.x}]   \nonumber \\
&+& \sin[{\it k_L} \,\ {\bf b_3 .x}] ) 
\eea
where, 
$\bf b_1 = - \frac{1}{2} e_x - \frac{\sqrt{3}}{2} e_y ; b_2 = e_x; b_3 = -\frac{1}{2} e_x + \frac{\sqrt{3}}{2} e_y$ and $\rm k_L = 2 \sqrt{3} \pi/\lambda_L$ ($\lambda_L$ is the laser wavelength and is 830 nm for the experiment under discussion). With these considerations $V_{\rm lat}$ is the height of the barrier between neighboring sites. The difference between the maximum and minimum values of $V_{\rm hex} ({\bf r})$ is 8 $V_{\rm lat}$.\\

The experiment uses a separate set of lasers to provide strong confinement in the third dimension, $V_{\rm conf}({\bf r})$:
\be
 V_{\rm conf}({\bf r}) = V_{\rm 1D} \,\  {\cos[\frac{2 \pi}{\lambda_{\rm 1D}} {\rm z}]} \approx \frac{V_{\rm 1D}}{2}(\frac{2 \pi}{\lambda_{\rm 1D}} )^2 {\rm z}^2 .
 \ee
This potential restricts the dynamics to two dimensions and we may take the wavefunction of the BEC in the third direction to be constant and Gaussian. Then the energy can be written as :\\

\bea
E_{\rm 2D} &=& \int d^2{\bf r} \sum_{\sigma=1,2} [-\frac{\hbar^2}{2m}\nabla^2 \psi_\sigma({\bf r}) + V_\sigma({\bf r}) |\psi_i({\bf r})|^2  \nonumber\\
&+& \frac{U_{\rm 2D}}{2}|\psi_\sigma({\bf r})|^4] + W_{\rm 2D} |\psi_1({\bf r})|^2  |\psi_2({\bf r})|^2    
\label{Energy2D}
\eea
where $U_{\rm 2D} =U_{\rm 3D}\sqrt{\frac{\sqrt{m V_{\rm 1D}}\,\ 2 \pi}{\lambda _{\rm 1D}\,\ h}}$ and $W_{\rm 2D} = W_{\rm 3D}\sqrt{\frac{\sqrt{m V_{\rm 1D}}\,\ 2 \pi}{\lambda _{\rm 1D} \,\ h}}$. In the experiment \cite{Sengstocktsfnsf}, $\lambda _{\rm 1D} = \lambda _{L} =$ 830 nm and $V_{\rm 1D} = 8.8 E_{\rm R}$. For these parameters, the weakest lattice yielding a Mott state is $V_{\rm lat} \approx 3.5 \,\ E_{\rm R}$ for two particles per unit cell within the Gutzwiller mean field approximation \cite{Sengstockhoneycomb}. \\

We assume a form of $\psi_1 (\bf r)$ and $\psi_2 (\bf r)$ which is consistent with the time-of-flight measurements :
\bea
\psi_1({\bf r}) &=& \sum_k \psi_1({\bf k}) \exp(- i\,\ \bf{k.r}) ,\\
\psi_2 ({\bf r}) &=& \sum_k \psi_2({\bf k}) \exp(- i\,\ \bf{k.r}) .
\eea
where $\bf k$ are the reciprocal lattice vectors of a honeycomb lattice. We insert this variational ansatz into eq.(\ref{Energy}) and minimize the energy with respect to the set of variational parameters $\psi_1({\bf k})$ and $\psi_2({\bf k})$. We find from our simulations that for all experimental parameters $\psi_1({\bf k}) = \psi_2^{*}({\bf k})$, where $\psi_2^{*}({\bf k})$ is the complex conjugate of $\psi_2({\bf k})$. This result is sensible and implies $\psi_1$ and $\psi_2$ are related by a lattice translation. \\

We perform the variational minimization in Fourier space rather than real space (where such minimization is usually done). This is equivalent to solving the Gross-Pitaevskii equation in real space within a single unit cell with periodic boundary conditions. Computationally, we find momentum space to be more efficient. Moreover, the experimental probes are all in momentum space. Similar approaches have been used by other authors \cite{Muellernld, Wuhoneycomb, dassarmaloop}.\\

\section{Method}
In k-space, the energy, eq.(\ref{Energy2D}) becomes :
\bea
\frac{E_{\rm 2D}}{E_{\rm R}} &=& \sum_{\{{\bf k,k_1,k_2,k_3}\} \epsilon \overline{\cal{L}}} \sum_{i=1,2} [3 \,\ k^2\psi_i^{*}({\bf k}) \psi_i({\bf k}) \nonumber  \\ 
&+&V_i({\bf k_1}) \psi_i^{*}({\bf k_2})  \psi_i({\bf k_2-k_1})  \nonumber\\
&+& \frac{U}{2} \psi_i^{*}({\bf k_1}) \psi_i^{*}({\bf k_2}) \psi_i({\bf k_3}) \psi_i({\bf k_1+k_2-k_3}) ]  \nonumber\\
&+& W \psi_1^{*}({\bf k_1}) \psi_1({\bf k_2}) \psi_2^{*}({\bf k_3}) \psi_2({\bf k_1+k_3-k_2}), \nonumber \\
\label{Energykspace}
\eea
where $\overline{\cal{L}}$ stands for the reciprocal lattice i.e ${\bf k} = ( a_1 {\bf b_1} + a_2 {\bf b_2})$, $ a_1 \,\ {\rm and} \,\ a_2$ being integers and $k = |\bf k|$. One can also generate this lattice from one of $\bf b_1$, $\bf b_2$ and $\bf b_3$, all explicitly given following Eq.(7). All energies $(V_i,U\,\ {\rm and}\,\ W)$ are expressed in terms of $E_{\rm R}$. \\

While we carried out unrestricted minimizations, our results are best illustrated by considering an ansatz where the low momentum physics is characterized by 2 complex numbers $t$ and $z$. In particular, we take $\psi_1({\bf k}) = t$ and $\psi_2({\bf k})= z$ for $\bf k = \{{\bf b_1,\ \ b_2,\ \ b_3} \}$ and $\psi_1({\bf k}) = z$ and $\psi_2({\bf k})= t$ for $\bf k = \{{\bf - b_1,\ \ - b_2,\ \ - b_3 }\}$. In terms of their real and imaginary parts, we write
\bea
t &=&  t_{\rm r} +{\it i} \,\  t_{\rm i} \,\  \rm and  \\
z &=& z_{\rm r} +{\it i} \,\  z_{\rm i} .
\eea
As has been mentioned in Section 1.B, the order parameter (OP) for the twisted superfluid state is  given by:
\be
OP = |\frac{|z|^2-|t|^2}{|z|^2+|t|^2}|
\ee

For our minimization, we restrict ourselves to $\rm |{\bf k}| \le 6$ giving us 159 complex variational parameters. We find that there are no differences if we use $\rm |{\bf k}| \le 4$ instead. Therefore, we believe our results faithfully reflect what would be found if an infinite number of  Brillouin zones were included. We gain further confidence in the convergence of our results by noting that the fraction of population occupying the $\rm |{\bf k}|  = 4 $ state when $U=0.05 E_{\rm R}$ and $V_{\rm lat} = 3.8 E_{\rm R}$ is about $0.0001\%$. It should also be noted that in the absence of interactions, at $V_{\rm lat} = 4 E_{\rm R}$, the real space Wannier functions have width $\frac{1}{k_L} \sqrt{\frac{2}{3}} $ and the probability of having $\rm |{\bf k}| \ge 2$ is less than 2 $\%$. Interactions tend to spread out the wavefunction, further reducing the occupation of high $|\bf k|$ states. In our simulations, we vary $U$ in the range $0.03 E_{\rm R}$ to $0.2 E_{\rm R}$ corresponding to various strengths of the transverse confinement. For the experiment, $U \approx 0.05 E_{\rm R}$. We also vary $\alpha$ in the range 0.08 to 0.3, corresponding to varying amounts of detuning of the laser beams.\\

\section{Results}
We do not find any evidence for the existence of the Twisted Superfluid state despite an extensive search of the parameter space. Since Eq.(\ref{Energykspace})  is a quartic form, it will in general have multiple minima and a number of other stationary points. The most grave concern with our results is that we might not have found the global minimum. To some extent, we can alleviate this concern by noting that the experiment finds a continuous symmetry breaking as a function of lattice depth. It therefore suffices to establish that our solution is a dynamically stable local minimum which is continuously connected to the symmetry-unbroken ground state at $V_{\rm lat}=0$. \\

\subsection{Local Energetic Stability}
\begin{figure}
\begin{center}
\includegraphics[scale=0.258]{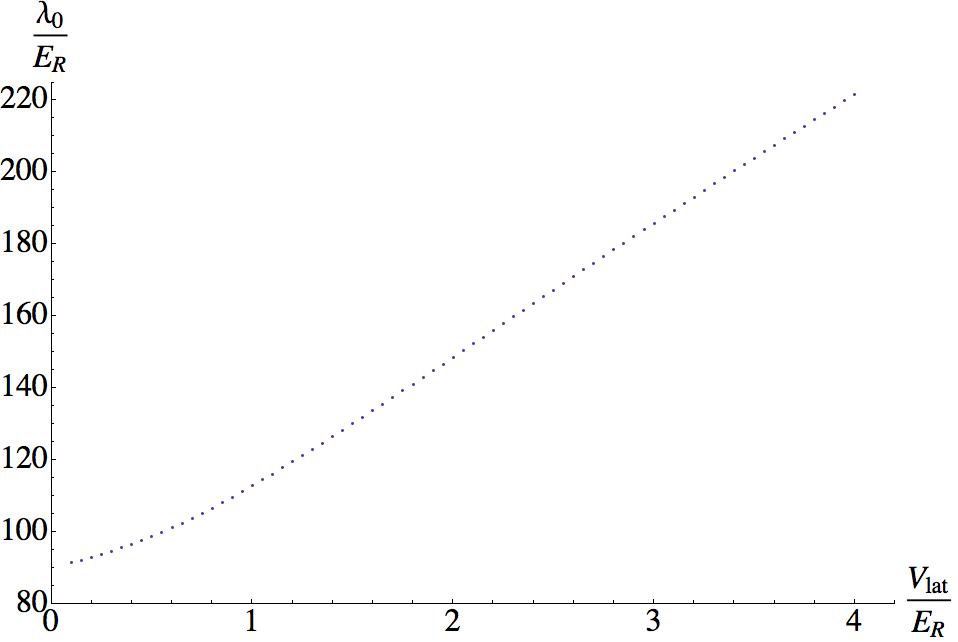}
\end{center}
\caption{Minimum eigenvalue of the Hessian, $\lambda_0$ in the Normal superfluid phase plotted against the lattice depth, $V_{\rm lat}$ (in units of $E_{\rm R}$) when $U=0.05 E_{\rm R}$ and 5 particles (of each species) are present per unit cell. All the eigenvalues of the Hessian are positive, thereby showing the stability of the normal phase. We conclude that there is no Twisted superfluid state for these potential depths. This result is illustrative of all parameter ranges we explored.} 
\label{Hessian}
\end{figure}

We check whether whether we have found a true minimum by looking at the eigenvalues of the Hessian $H$ defined by :
\be
H_{ij} =  \frac{\partial^2 E}{\partial a_i\partial a_j} ,
\ee
where $a_i$ and $a_j$ are real variational parameters (corresponding to the real and imaginary parts of $\psi(\bf k)$). We find that for all parameters, the eigenvalues of $H$ are positive. This implies that we have at least found a local minimum. In Figure \ref{Hessian}, we plot the minimum eigenvalues of the Hessian for different values of the lattice depth ($V_{\rm lat})$ at the illustrative interaction strength, $U=0.05 E_{\rm R}$ and $\alpha = 0.14$, for five particles (of each species) per unit cell. \\

We further illustrate the stability of our theory by doing two separate numerical experiments : \\

\begin{figure}
\begin{center}
\includegraphics[scale=0.35]{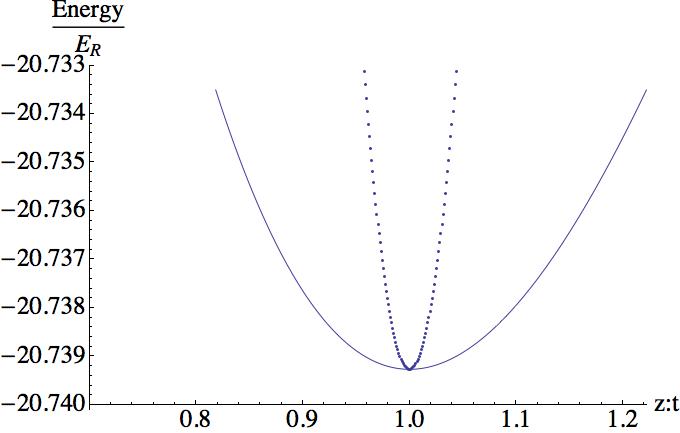}
\end{center}
\caption{Slice through the energy landscape at $V_{\rm lat} = 1.8 E_{\rm R}$ and $U = 0.05 E_{\rm R}$ and 5 particles (of each species) per unit cell. 
Dotted curve: The ratio Re[z]:Re[t]is varied and the energy is found  by minimizing with respect to the other variational parameters. 
Solid curve: Same, but with varying Im[z]:Im[t]. 
We find that the overall energy minimum occurs when Re[z] = Re[t] and Im[z] = Im[t].}
\label{trzr}
\end{figure}

(a) Fix the ratio of ${ z_{\rm r} \,\ ({\rm Re}[z])\,\  {\rm to} \,\ t_{\rm r} \,\ ({\rm Re}[t])}$  and vary the remaining variational parameters to find the energy minima. We find that the minimum of the energy occurs when ${z_{\rm r} : t_{\rm r }=1}$ and there are no other local minima. The dotted curve shows this in Figure \ref{trzr}. \\

(b) Fix the ratio of ${ z_{\rm i} \,\ ({\rm Im} [z]) \,\ {\rm to} \,\ t_{\rm i} \,\ ({\rm Im}[t])}$ and vary the remaining variational parameters to find the energy minima. We find that the minimum of the energy occurs when ${z_{\rm i} : t_{
\rm i}}$ =1 and there are no other local minima. The solid curve shows this in Figure \ref{trzr}.\\

We conclude that there is no second order phase transition  within mean field theory.

\subsection{Local Dynamic Stability}
We also check whether the minimum found is unstable against perturbations. This is done by looking at the Gross-Pitaevskii equation :
\be
i \hbar \frac{\partial \psi}{\partial t} = \frac{\partial E}{\partial \psi^{*}}
\ee
This would imply :
\be
i \hbar \frac{\partial \delta a_j}{\partial t} = \frac{\delta E}{\delta a_j}  \approx \sum_{l} \frac{\partial^2 E}{\partial a_j \partial a_l} \delta a_l
\ee
Taking the real and imaginary parts of both sides, we get the eigenvalue equations
\bea
\hbar \omega \,\ u = M u
\eea
where, 
 \[ M = \left[
 \begin{array}{cc}
$Re$[H] & $-Im$[H] \\
$Im$[H] & $Re$[H] 
\end{array} \right]\]

We look at the eigenvalues of this matrix, $M$. A complex eigenvalue would signify the presence of a mode which will grow with time, thus rendering this ground state unstable. We find that all the eigenvalues are real. Thus, the minimum that we have found is also dynamically stable. This is a generic feature of quantum systems: Energetic stability implies dynamic stability \cite{PethickSmith}.\\

\section{Discussion}
Given that our mean-field treatment of Eq. (5) fails to reproduce the experimental observations, we must now confront the question of what additional physics is needed to produce a twisted superfluid state. In this section, we present a tight-binding model which has a twisted superfluid ground state and discuss connections with our approach. Namely, consider a Hamiltonian:
\be
H =  \sum_{<\rm ij>} \left( -t (\hat{a}_{\rm i \uparrow}^{\dagger} \hat{a}_{\rm j\uparrow} + \hat{a}_{\rm i\downarrow}^{\dagger} \hat{a}_{\rm i\downarrow}) +  t_{\rm cf} (\hat{a}_{\rm i \uparrow}^{\dagger} \hat{a}_{\rm j \downarrow}^{\dagger}  \hat{a}_{\rm j \uparrow} \hat{a}_{\rm i\downarrow}) + h.c. \right).
\label{nn1}
\ee
Here, $\rm {a}_{i \sigma}$ annihilates a particle labelled by the spin index $\sigma$ on site i, and the sum is over all nearest neighbor sites of a honeycomb lattice. The parameters $t$ and $t_{\rm cf}$ represent single particle and counter-flow hopping. We consider a mean-field ansatz where  $\hat{a}_{\rm j \sigma}$ is replaced by a c-number, which can take one of two values, depending on which sub-lattice site j belongs to (see Fig. 1):
\bea 
a_{\rm j\uparrow} = \sqrt{n_{+}} \,\ \exp (-{\it  i}\,\ \delta/2) \,\,\,\,\ \rm sublattice\,\,\ A\\
a_{\rm j\uparrow} = \sqrt{n_{-}} \,\ \exp (+ {\it i}\,\ \delta/2) \,\,\,\,\ \rm sublattice\,\,\  B
\eea
and
\bea
a_{\rm j\downarrow} = \sqrt{n_{-}} \,\ \exp (+{\it  i}\,\ \delta/2) \,\,\,\,\ \rm sublattice\,\,\ A\\
a_{\rm j\downarrow} = \sqrt{n_{+}} \,\ \exp (- {\it i}\,\ \delta/2) \,\,\,\,\ \rm sublattice\,\,\ B
\eea

A twisted superfluid corresponds to $\delta \ne 0$ and physically can be interpreted as a state where there are microscopic single particle single particle currents, which are precisely balanced by microscopic counterflow currents. The mean-field energy per site is :
\be
E = \left(-12 t \sqrt{n_{+} n_{-}} {\cos({\delta})} + 6 t_{\rm cf}\,\ n_{+} n_{-} {\cos(2 {\delta})} \right).
\label{energ}
\ee
The lowest energy state has $\delta \ne 0$ if :
\be
2 t_{\rm cf} (n_{+} n_{-}) > t \sqrt{n_{+} n_{-}}
\label{ineq}
\ee

Our model in Eq. (5) contains terms of the form as those in Eq. (\ref{energ}). For deep lattices \cite{Blochrev},
\be
t \sim |a|^{-3/2}\,\,\ {\exp}(-\pi\sqrt{V_{\rm lat}/E_{\rm R}}/2)
\ee
and
\be
t_{\rm cf} \sim |a|^{-3}\,\,\ {\exp}(-\pi\sqrt{V_{\rm lat}/E_{\rm R}}),
\ee
where a is the distance between nearest neighbors. The exponential suppression of $t_{\rm cf}$ means that for any reasonable particle density, Eq.(\ref{ineq}) is not satisfied. On the other hand, quantum fluctuations suppress single particle hopping more than counterflow {\cite{Svistunovscf1, Svistunovscf2, Demlernjp2003,Clarkpra2009,Kawashimapra2011}, and a beyond mean field theory treatment of Eq.(5) could yield a twisted superfluid. Thus, the observations of Soltan-Panahi {\it et al.} \cite{Sengstocktsfnsf} may be evidence of non-mean field physics.
\section*{Acknowledgements}

We would like to thank Yariv Yanay and Mukund Vengalattore for critical comments on the manuscript. This paper is based on work supported by the National Science Foundation under Grant no. PHY-1068165.


\begin{thebibliography}{10}

\bibitem{Sengstocktsfnsf} P. Soltan-Panahi, D.-S. L\"{u}hmann, J. Struck, P. Windpassinger and K. Sengstock, Nature Physics {\bf 8}, {71-75} (2012).

\bibitem{SengstockFrustratedScience2011} J. Struck, C. \"{O}lschl\"{a}ger, R. Le Targat, P. Soltan-Panahi, A. Eckardt, M. Lewenstein, P. Windpassinger and K. Sengstock, \SC{333}{996}{2011}.

\bibitem{Hemmerichpwave2011} G. Wirth, M. \"{O}lschl\"{a}ger and A. Hemmerich, Nature Physics {\bf 7}, {147-153} (2011).

\bibitem{Hemmerichfwave2011} M. \"{O}lschl\"{a}ger,  G. Wirth and A. Hemmerich, \PRL {106}{015302}{2011}.

\bibitem{KupferschmidtMueller} J. N. Kupferschmidt  and E. J. Mueller, \PRA{82}{023618}{2010}.

\bibitem{Sengstockhoneycomb} P. Soltan-Panahi, J. Struck, P. Hauke, A. Bick, W. Plenkers, G. Meineke, C. Becker, P. Windpassinger, M. Lewenstein and Klaus Sengstock, Nature Physics {\bf 7}, {434-440} (2011).

\bibitem{Muellernld} E. J. Mueller, Phys. Rev. A {\bf 66}, 063603 (2002)

\bibitem{Wuhoneycomb} Z. Chen and B. Wu,  \PRL{107}{065301}{2011}.

\bibitem{dassarmaloop} H-Y Hui, R. Barnett, J. V. Porto and S. Das Sarma, \PRA{86}{063636}{2012}.

\bibitem{PethickSmith} C. J. Pethick and H. Smith, {\it Bose-Einstein Condensation in Dilute Gases}. Cambridge University Press, Cambridge (2002).

\bibitem{Blochrev} I. Bloch, J. Dalibard and W. Zwerger,  \RMP {80} {885} {2008}.

\bibitem{Svistunovscf1} A. B. Kuklov and B.V. Svistunov, \PRL{90}{100401}{2003}.

\bibitem{Svistunovscf2} A. B. Kuklov, Nikolay Prokof'ev and B.V. Svistunov, \PRL{92}{050402}{2004}.

\bibitem{Demlernjp2003} E. Altman, W. Hofstetter, E. Demler and M. D. Lukin, \NJP{5}{113}{2003}

\bibitem{Clarkpra2009} A. Hu, L. Mathey, I. Danshita, E. Tiesinga, C. J. Williams, C. W. Clark, Phys. Rev. A {\bf 80}, 023619 (2009).

\bibitem{Kawashimapra2011}T. Ohgoe and N. Kawashima, Phys. Rev. A {\bf 83}, 023622 (2011).










\end{thebibliography}
\end{document}